\documentclass[12pt]{revtex4}
\usepackage{graphics,graphicx,amssymb}
\begin{document}

{\tiny{\bf{for the Proceedings of the International Workshop on Statistical-Mechanical Informatics
September 14–17, 2008, Sendai, Japan}}}

\title{Non-classical Role of Potential Energy in Adiabatic Quantum
Annealing}

\author{Arnab Das}

\affiliation{The Abdus Salam International Center for Theoretical Physics,
Starda Costiera 11, 34014 Trieste, Italy}

\email{arnabdas@ictp.it}

\begin{abstract}
Adiabatic quantum annealing 
is a paradigm of 
analog quantum computation, where 
a given computational
job is converted to the task of finding the global minimum
of some classical potential energy function and 
the search for the global potential 
minimum is performed by employing external 
%means of
%quantum annealing dynamics, which
%consists of the employment of external
%(i.e., the profile of the 
%potential energy with respect to the configurations of the 
%optimizing variables or degrees of freedom) using 
kinetic quantum fluctuations and subsequent slow reduction (annealing) 
of them. 
In this method, the entire potential energy landscape (PEL)
may be accessed simultaneously through
a delocalized wave-function, 
in contrast to a classical search, where
the searcher has to visit different 
points in the landscape (i.e., individual classical configurations) 
sequentially. Thus in such searches, the role
of the potential energy might be significantly 
different in the two cases. Here we discuss this in the context of 
searching of a single isolated hole (potential minimum)
in a golf-course type gradient free PEL. We show, 
that the quantum particle would be able to locate 
the hole faster if the hole is deeper, while
the classical particle of course would have no scope to 
exploit the depth of the hole. 
We also discuss the effect of
the underlying quantum phase 
transition on the adiabatic dynamics.         
\end{abstract}

\maketitle

\section{Introduction}
Adiabatic quantum annealing (AQA) 
\cite{BC:Kado}-\cite{Somma} is a method of finding the
ground state (minimum energy state)
of a given classical Hamiltonian by employing external
quantum fluctuations and subsequent adiabatic reduction of them.
One is given with a classical Hamiltonian $\mathcal{H}$, which
may be a physical Hamiltonian with many degrees of freedom,
or a suitable mathematical function depending on many variables, 
and the task is to determine its global minimum.
In order to introduce the quantum fluctuations
necessary for the AQA of such a Hamiltonian, one adds a quantum
kinetic part $\mathcal{H}^{\prime}(t)$ to it,
such that $\mathcal{H}^{\prime}(t)$ and $\mathcal{H}$ do not commute.
Initially, one keeps 
$|\mathcal{H}^{\prime}(t=0)| \gg |\mathcal{H}|$ so that
the total 
Hamiltonian 
$\mathcal{H}_{tot}(t) = \mathcal{H}^{\prime}(t) + \mathcal{H}$ 
is well approximated by the kinetic part only
($\mathcal{H}_{tot}(0) \approx \mathcal{H}^{\prime}(0)$).
If the system is initially prepared to be in the
ground state of $\mathcal{H}^{\prime}(0)$
(one chooses $\mathcal{H}^{\prime}(0)$ to have a
easily realizable ground state) and
%Now since $\mathcal{H}_{tot}(0) \approx \mathcal{H}^{\prime}(0)$, the
%overlap $|\langle \psi (t)|E_{-}(t)\rangle|$ 
%between the lowest eigenvalue state
%(we will call it instantaneous ground state) $|E_{-}(t)\rangle$
%of the total Hamiltonian $\mathcal{H}_{tot}(t)$ and the instantaneous state
%$|\psi(t)\rangle$ of the evolving system will be close to unity at $t = 0.$
%(since $|\psi (0)\rangle$ is taken to be the ground state of
%$\mathcal{H}^{\prime}(0)$).
$\mathcal{H}^{\prime}(t)$ is reduced slowly enough,
then according to
the adiabatic theorem of quantum mechanics, the overlap
$|\langle \psi (t)|E_{-}(t)\rangle|$, 
between the instantaneous lowest-eigenvalue state $|E_{-}(t)\rangle$
and the instantaneous state of $|\psi(t)\rangle$ of the evolving system,
will always stay near 
its initial value (which is close to unity, since 
$\mathcal{H}_{tot}(0) \approx \mathcal{H}^{\prime}(0)$). Hence
at the end of such an evolution, when $\mathcal{H}^{\prime}(t)$ is reduced to
zero at $t = \tau$ (the annealing time), 
the system will be found in a state $|\psi(\tau)\rangle$
with
$|\langle \psi (\tau)|E_{-}(\tau)\rangle| \approx 1$,
where $|E_{-}(\tau)\rangle$
is the ground state of $\mathcal{H}_{tot}(\tau)$,
which is nothing but the surviving
classical part $\mathcal{H}$. Thus
at the end of an adiabatic annealing the system is found in the ground
state of the classical Hamiltonian with a high probability.
Based on this principle, algorithms can be framed to anneal
complex physical systems like spin glasses
as well as the objective functions of hard combinatorial
optimization problems (like the Traveling Salesman Problem),
towards their ground (optimal) states 
\cite{Farhi, Santoro-Sc, Santoro-TSP, AD-ZTQMC}. 

%However, in order to have the probability of reaching the ground state
%exactly unity, one needs to be able to
%tune both $\mathcal{H}$ and $\mathcal{H}^{\prime}$, so that 
%$\mathcal{H}_{tot}(0) = \mathcal{H}^{\prime}(0)$, 
%$\mathcal{H}_{tot}(\tau) = \mathcal{H}(\tau)$ and the evolution
In order to ensure adiabaticity, the evolution should be such
that
\begin{equation}
\tau \gg \alpha; \quad
\alpha = \frac{
|\langle \dot{\mathcal{H}_{tot}}
\rangle|_{max}}{\Delta^2_{min}},
\label{adia-cond}
\end{equation}
\noindent
where
\begin{eqnarray}
|\langle\dot{\mathcal{H}_{tot}}\rangle|_{max} &=&
\max_{0 \le s\le 1}\left[
\left|\left\langle
E_{-}(s)\left|\frac{d\mathcal{H}_{tot}}{ds}\right|E_{+}(s)
\right\rangle\right|\right] \nonumber \\
\Delta^2_{min} &=& \min_{0 \le s \le 1}\left[ \Delta^2(s)\right];\quad
s = t/\tau; \quad 0\le s \le 1,
\label{adia-factors}
\end{eqnarray}
\noindent
$|E_{+}(t)\rangle$ being
the instantaneous first excited state of the total
Hamiltonian $\mathcal{H}_{tot}(t)$, $\Delta(t)$ being the instantaneous gap
between the ground state and the first excited state energies and $\alpha$
being the adiabatic factor (for a simple proof
see \cite{Sarandy-Lidar}).

One key feature, believed to be behind the success of AQA over the
classical ones \cite{Farhi, Santoro-Sc, Santoro-TSP} in glass-like
rugged PEL, is the ability of the quantum systems
to tunnel easily through potential energy barriers even if they are 
very high, provided they are narrow enough, in contrast to the
the classical ones, which always has to scale the barrier 
height with its kinetic 
energy (temperature) irrespective of the width \cite{BC:adbk, Brook,
AD-pap, Farhi-SA-QA, BKC-pap}. 
Here we show that there is another
aspect which makes a quantum mechanical searcher more advantagous over the
classical ones- it can utilize the depth of the potential energy minimum
in locating it in absence of any potential gradient which a classical
searcher cannot.

\section{Searching a hole on a gradient-free PEL}

We consider a lattice with $N$ sites, $|i\rangle$
denoting the state of a particle localized at the $i$-th site. 
At each site, there is a potential, which is
 zero at all the sites $i \ne w$, 
and is $-\chi$ at $i = w $, where $w$ is chosen randomly. 
Thus the PEL is essentially a flat one
without any gradient, with a single hole (minimum) at $i = w$ with a depth
$\chi$. This is precisely some kind of analog version of Grover's algorithm
for searching a particular entry in an unstructured database 
\cite{Grover} - \cite{Roland-Cerf}. But in those studies, the possibility
of utilizing the depth of the hole in favor of faster search was
not considered, and the gain over the classical ones is limited by
the optimal Grover's bound of $\mathcal{O}(\sqrt{N})$ speed-up 
\cite{Roland-Cerf, Mosca}.
 
Let us consider that the lattice points are connected to each other
by an infinite range hopping term $\Gamma$ between any
two sites. The question is how fast a particle can locate the
hole starting from a state which does not assume any knowledge of the
position of the hole, by reducing its kinetic energy $\Gamma$, and tuning
the hole depth $\chi$.   
The Hamiltonian for a particle on such a lattice will be
given by-
%To do quantum annealing, we evolve $\chi(t)$ from zero to its
%final value $\chi_{0}$ keeping $\Gamma$ fixed at some moderate constant value.
%The total Hamiltonian is given by
%$\mathcal{H}_{tot} = \mathcal{H}_c + \mathcal{H}$
%with $\mathcal{H}_C = -chi_{0}|w\rangle\langlew|$ and $\mathcal{H}_{kin} = $
\begin{equation}
\mathcal{H}_{tot}(t) = -\chi(t)|w\rangle\langle w|
-\Gamma(t)\sum_{i,j; i\ne j}|i\rangle\langle j|; 
\quad \chi(t), \Gamma(t) > 0.
\label{ad-ham}
\end{equation}
\noindent 
In order to anneal the particle to the hole, 
one has to reduce $\Gamma$ from a very high value to a very low
final value and tune $\chi$ in the opposite manner, so that 
$\Gamma(t=0) \gg \chi(t=0)$ and $\Gamma(t=\tau) \ll \chi(t=\tau)$, where
$\tau$ is the annealing time. The
evolution should satisfy the adiabatic condition (\ref{adia-cond}). 
%In order to investigate the condition of adiabatic annealing of the system,
%we note that the eigenvalue problem for the system can be exactly solved.
The eigen-spectrum of $\mathcal{H}_{tot}(t)$ 
consists of a ground state $|E_{-}(t)\rangle$
and first excited state $|E_{+}(t)\rangle$ 
(in the order of 
increasing eigen values)
with energies
\begin{equation}
E_{\pm}(t) = -\frac{1}{2}\left[(N-2)\Gamma +\chi
\pm \sqrt{(N\Gamma - \chi)^{2} + 4\chi\Gamma}\right] 
\label{eigenvalues}
\end{equation}
\noindent
respectively, all the time 
dependencies being implicit, through the time dependence of
$\Gamma$ and $\chi$. The instantaneous gap is thus given by
\begin{equation}
\Delta(t) = \left|\sqrt{(N\Gamma - \chi)^{2} + 4\chi\Gamma}\right|.
\label{gap}
\end{equation}
\noindent
The instantaneous first and second excited states 
$|E_{\pm}(t)\rangle$ are given by -
\begin{equation}
|E_{\pm}(t)\rangle 
= \frac{1}{\sqrt{C_{\pm}^{2}(t) + N - 1}}\left[C_{\pm}(t)|w\rangle 
+ \sum_{i\ne w}^{N} |i\rangle\right], 
\label{eigenstts}
\end{equation}
\noindent
where 
\begin{equation}
C_{\pm}(t) = \frac{1}{2\Gamma}\left[-(N - 2)\Gamma + \chi \mp \Delta \right] 
%S_{\pm}(t) &=& \sqrt{C_{\pm}^{2}(t) + N - 1}
\label{coeffs}
\end{equation}
\noindent
The second excited state is $(N-2)$-fold degenerate, with eigenvalue $-\Gamma$,
and the time evolving Hamiltonian never mixes the first two eigenstate
with any of the second excited states. This can be easily
argued noting that a state of the form
$|E_{2}\rangle = \frac{1}{\sqrt{2}}(|i\rangle - |j\rangle)$
$(i,j \ne w)$ is an eigenstate of $\mathcal{H}_{tot}(t)$ with
eigenvalue 
$-\Gamma$, and $\langle E_{2}|E_{-}(t)\rangle 
= \langle E_{2}|E_{+}(t)\rangle = 0$
for all $t$. For all allowed combinations of $i$ and $j$ we get $(N-2)$ such
linearly-independent eigenstates. Form these $(N-2)$ 
linearly-independent eigenstates
we can construct $(N-2)$
mutually orthogonal eigenstates, 
each of which will obviously satisfy the above
non-mixing condition. Thus we have to take care of only two 
lowest lying states and the gap between them.
%Here one may note that if $\chi$ is independent of $N$, 
%the gap $\Delta$ 
%scales approximately linearly with $N$ 
%for large $N$ at any given
%non-zero value of $\Gamma$, and so does the term 
%$|\langle\dot{\mathcal{H}_{tot}}\rangle|$. 
%But at $\Gamma = 0$, the $N$-dependence of the former 
%vanishes, but not the latter, if $\Gamma$ has a linear 
%time dependence. Thus
%one needs a linear $N$-dependence in $\chi$ in order to obtain an 
%adiabatic factor $\alpha$ that does not diverge with $N$,
%as $N \rightarrow \infty$.
Henceforth we will consider
only the large $N$ limits, and replace $'='$ by 
$'\approx'$ when the correction
will vanish in the said limit. 

%Physically this means, 
%the depth of the hole is to be increased in order
%to get it sensed and successfully tracked by the
%searching wave-function if the space of search is made larger. Such 
%a non-local sensing of the hole-depth and utilizing it for
%tracking the hole is impossible for a classical searcher, 
%since it cannot sense the hole depth
%unless it drops right into it.  

In investigating the condition of adiabatic 
evolution we consider two separate cases. In general, in a
quantum annealing program, one might not have the facility of 
tuning both the potential 
energy part and the 
kinetic part in practice, since, say, one might 
depend on the strength of interactions between the
elementary constituents (hard to tune), while the 
other might be introduced through an applied external
field (easily tunable). Hence in our analysis we 
have considered two separate cases - in the first case
we tune $\chi$ keeping $\Gamma$ constant, while in 
the second case we do the reverse. 

\subsection{Constant-$\Gamma$ Annealing}  

\begin{figure}
\centerline{
\resizebox{10.0cm}{!}{\rotatebox{0}{\includegraphics{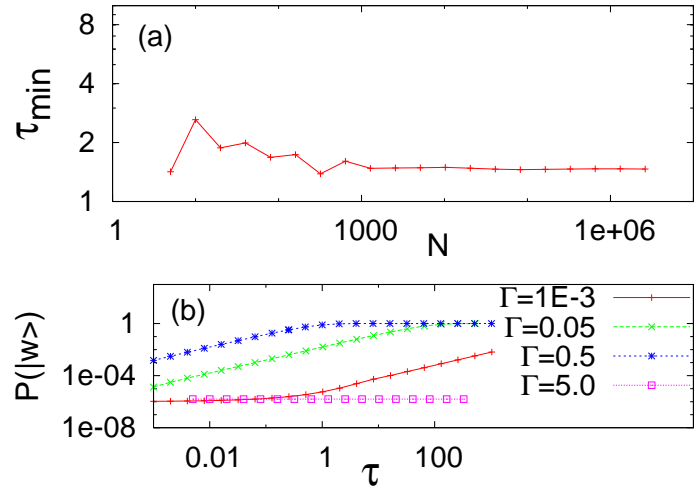}}}}
\caption{\footnotesize{Panel (a) in the figure
shows the variation of the
minimum time $\tau_{min}$ required
to achieve the target success probability $P_{T} = 0.33$
for the constant-$\Gamma$ annealing 
($\Gamma(t) = \Gamma_{0}$, $\chi(t) = \chi t/\tau$; 
$r = 2.0$, $\Gamma_{0} = 0.5$), obtained by solving
the time-dependent Schr\"{o}dinger equation numerically. 
Panel (b) shows the variation of final probability $P(|w\rangle)$ of finding
the system in state $|w\rangle$, with annealing time $\tau$, 
for different final 
value of $\Gamma$. Here we have taken
$N = 10^6$.}}
\label{qan}
\end{figure}

We start with the initial state 
$|\psi(0)\rangle =\frac{1}{\sqrt{N}} \sum_{i}^{N}|i\rangle$ (which is
of course the ground state of $\mathcal{H}_{tot}(0)$)
and adopt a linear annealing schedule: 
%with an explicit
%$N$-dependence of $\chi$ -
\begin{equation}
\Gamma = \Gamma_{0}; \quad \chi = \chi_{0}\frac{t}{\tau}; \quad \chi_{0} 
= rN\Gamma_{0}.
\label{Const-G-schdl}
\end{equation}
Here $r$ is an arbitrary factor which might also depend on $N$.
When $\Gamma$ is kept constant, we have to keep it 
sufficiently low (or, in other words, $\chi_{0}$ sufficiently high), 
so that the
ground state of ${\mathcal{H}}_{tot}(\tau)$ has a substantial 
(non-vanishing in the $N \rightarrow \infty$ limit) overlap 
with $|w\rangle$. At $t = \tau$, we get from Eq. (\ref{coeffs}),
the amplitude of $|w\rangle$ in the final ground state to be 
$C_{-}(\tau)/\sqrt{C_{-}^{2}(\tau) + N - 1}$,
where 
\begin{eqnarray}
C_{-}(\tau) &=& 1 + \frac{N(r - 1)}{2} 
              + \frac{\left|\sqrt{N^{2}(r - 1)^{2} + 4rN}\right|}{2}.  
%\nonumber \\
%&\approx&  1 + \frac{N(r - 1)}{2} + \frac{|N(r - 1)|}{2}.    
\label{ConstG-final-ovlp}
\end{eqnarray}  
\noindent Clearly, if $r > 1$, then the 
amplitude $C_{-}(\tau) \gtrsim N$, and thus the overlap amplitude 
$\sim {\mathcal{O}}(1)$, whereas if $r < 1$, then $C_{-}(\tau) \sim 1$, 
and the amplitude vanishes as 
$N \rightarrow \infty$. Thus to be able to locate the hole 
at the end, we have to take $r > 1$. In fact,  
The adiabatic factor in that case ($r > 1$) is given by 
\begin{equation}
\alpha_{1} =
\frac{|\langle\dot{\mathcal{H}_{tot}}(t)\rangle|_{max}}
{|\Delta^{2}(t)|_{min}} =
 \left(\frac{r\Gamma_{0}^{2}}{\tau}\right)
\frac{N\sqrt{N-1}}{|\Delta^{3}(t)|_{min}}
\label{ConstG-af-t}
\end{equation}
\noindent This has its maximum at 
$t_{m} = \tau(N-2)\Gamma_{0}/\chi_{0} \approx \tau/r $, and the maximum 
value is given by 
\begin{equation}
\alpha_{1} = \frac{\chi\sqrt{N}}{8\Gamma_{0}^{2}(N-2)^{3/2}} 
\approx \frac{r}{8\Gamma_{0}}.
\label{ConstG-af-max}
\end{equation}
\noindent This clearly shows that if the hole-depth $\chi_{0}$ scales 
linearly with the size $N$ of the search space 
(i.e., $r$ is independent of $N$), 
the search can be completed in a time
independent of $N$. We calculated numerically the 
minimum time $\tau_{min}$
required for obtaining a targeted succes
probability 
$P_{T} = |\langle \psi(\tau)|w\rangle|^{2} = 0.33$ 
%($|\psi(\tau)\rangle$ being the final wave function) 
for different $N$, through many
decades. The evolution is computed solving
time-dependent Schr\"{o}dinger equation numerically and
 $\tau_{min}$ is figured out up to an accuracy of $10^{-4}$ employing
the following bisection scheme. We first figure out arbitrarily a high 
value of $\tau$ (call it $\tau_{hi}$) for which 
$P(|w\rangle) \equiv |\langle\psi(\tau)|w \rangle|^{2} > P_{T}$. Next
we find a low $\tau$ ($\tau_{lo}$), for which $P(|w\rangle) < P_{T}$.
Then we evaluate $P(|w\rangle)$ for $\tau = \tau_{m} = (\tau_{hi} + \tau_{lo})/2$.
If the result is greater than $P_T$, then we replace $\tau_{hi}$ by $\tau_{m}$
(and retain old $\tau_{lo}$), else we replace $\tau_{lo}$ by $\tau_{m}$
(and retain old $\tau_{hi}$) and repeat the same process. We go on iterating
until the the value of $|P(|w\rangle) - P_{T}|$ for both $\tau_{hi}$
and $\tau_{lo}$ lies within some desired accuracy limit.   
%(i.e., running for higher ($\tau_{high}$) and
%lower values($\tau_{low}$) of $\tau$ necessary
%for reaching the target probability, and then decreasing the difference
% between $\tau_{high}$ and $\tau_{low}$ by bisecting it.
The results (Fig. \ref{qan}a)
clearly show that $P(|w\rangle)$ becomes
independent of $N$ for large $N$, as expected.

The relaxation behavior for large $N$ for a given annealing time $\tau$
of course depends on the value of $\Gamma_{0}$ (see Fig. \ref{qan}b).
If $\Gamma_{0}$ is too small,
the system takes a longer time to feel the changes in the landscape,
and hence the adiabatic
relaxation requires longer time (the adiabatic factor becomes bigger;
see Eq. \ref{ConstG-af-max}). On the
other hand, if $\Gamma_{0}$ is too large, the ground state itself is
pretty delocalized,
and hence the final state, though more closer to the ground state,
has again a small overlap with
the target state $|w\rangle$. For $\Gamma_{0} \approx 0.5$, the schedule 
is found to be optimal (Fig. \ref{qan}(b)).
The relaxation behavior is seen to be linear with the annealing time 
$\tau$ for large $\tau$.

\subsection{Constant-$\chi$ Annealing}

\begin{figure}[h]
\centerline{
\resizebox{8.0cm}{!}{\rotatebox{0}{\includegraphics{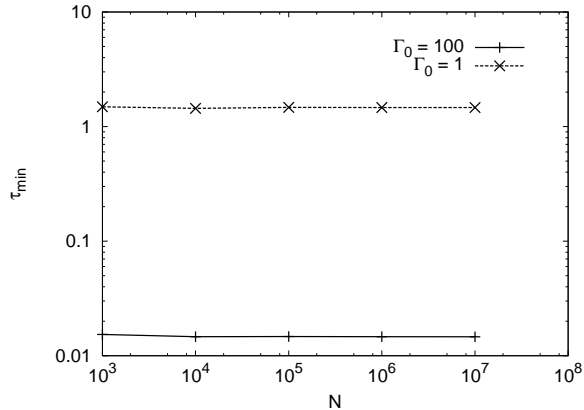}}}}
\caption{\footnotesize{The figure
shows variation of the
minimum time $\tau_{min}$ required
to achieve success probability $P_{T} = 0.9$, with $N$
for the constant-$\chi$ annealing. The result is obtained by solving
the time-dependent Schr\"{o}dinger equation numerically
with $\Gamma(t) = \Gamma_{0}(1 - t/\tau)$, $\chi(t) = \chi_{0} = r\Gamma_{0}N$
for $r = 0.5$.}}
%Initially, the system is delocalized
%equally over all sites and evolves with time
%according to time dependent Schr\"{o}dinger equation
%with the Hamiltonian (\ref{ad-ham}).
%As expected from exact analytical result for the adiabaticity condition,
%it is seen that $\tau_{min}$ becomes
%independent of $N$.}}
\label{ConstChi}
\end{figure} 

Next, we consider the case where $\chi$ is kept fixed and $\Gamma$ is
reduced linearly, i.e., $\chi = \chi_{0} = rN\Gamma_{0}$ and 
$\Gamma = \Gamma_{0}(1 - t/\tau)$. In this case, we start with the
same democratic initial state $|\psi_{0}\rangle$ as in the previous case, 
which is not of course the ground state in presence of the hole 
(we cannot construct the actual initial ground state without the explicit
knowledge of the location of the hole). All we need to show
in this case, is that it has
a non-vanishing (i.e., non-zero in the $N \rightarrow \infty$ limit) 
overlap $|\langle\psi(0)|E_{-}(0)\rangle|$ between our
initial state and the true ground state at $t = 0$. If we can assure 
adiabaticity for the subsequent evolution, this overlap will be conserved
and would emerge as the final overlap between 
$|\psi(\tau)\rangle$ and $|w\rangle$.
Calculation similar to that of the previous section shows 
that, for $r < 1$, $C_{-}(0) \rightarrow 0$ as 
$N \rightarrow \infty$, implying that the overlap actually 
tends to unity in that limit, while it 
vanishes for $r > 1$. 
The adiabatic factor in the former case $(r < 1)$ is given by 
\begin{equation}
\alpha_{2} \approx \frac{1}{8\Gamma_{0}r^{2}},
\label{ConstChi-af-max}
\end{equation}
\noindent which is again 
a constant independent of $N$, if $r$ is $N$-independent (i.e., 
$\chi_{0}$ is linear in $N$). 
We present a similar numerical
calculation, as in the previous section, for 
the variation of $\tau_{min}$ with $N$ in Fig. (\ref{ConstChi}).
It shows that the value of 
$\tau_{min}$ becomes independent of $N$ in the large-$N$
limit (its asymptotic value, however depends on $\Gamma_{0}$).

\section{The Underlying Transition}
Let us concentrate on the rather interesting case of $N$-independent $r$. 
The expression for the gap $\Delta$ (Eq. \ref{gap}) shows that
the gap scales linearly with $N$ everywhere except at the 
point $N\Gamma = \Gamma_{c} = \chi,$ 
where it scales as
$\sqrt{N}$. In our notation, this happens at the instant $t^{*} = \tau/r$ 
in the case of constant-$\Gamma$ annealing, and at $t^{**} = \tau(1 - r)$ 
in the
constant-$\chi$ case. For successful annealing
one cannot avoid this point during
the evolution in both cases. To see what happens at that point, 
let us focus on the amplitude $|\langle w | E_{-}(\chi,\Gamma)\rangle$
%of $|w\rangle$ in the ground state of $\mathcal{H}_{tot}(\chi,\Gamma)$, 
for $\Gamma > \Gamma_{c}$ and  $\Gamma < \Gamma_{c}$. 
Let us consider, say, the $\Gamma$-constant annealing case.
We have from
Eq. (\ref{coeffs}), $C_{-}(t) \approx (-N\Gamma_{0}(1 - rt/\tau) 
+ N\Gamma_{0}|1 - rt/\tau|)/2\Gamma_{0}$. 
Taking the modulus into account, one finds
$C_{-}(t) \approx 0$ for $t < t^{*}$ (which means the amplitude vanishes), 
while $C_{-}(t) \sim N$ for 
$t > t^{*}$ (the amplitude tends to unity). This means the ground state of
$\mathcal{H}_{tot}(\chi,\Gamma)$ 
undergoes a global change in character from a completely delocalized one to a 
completely localized one at this special point. Thus, 
to follow the ground state, a lot of tunneling from all the sites to the hole
is required at the point. Here the depth of the hole
plays a crucial role in making this tunneling possible 
preventing the gap $\Delta$ from closing at the transition point.
If however, the system passes this point very fast, this 
massive tunneling might remain incomplete and a resultant loss in the 
adiabaticity may occur. Similar argument holds for the
 $\chi$-constant annealing. 

If instead of the Hamiltonian we considered, one takes
a bounded version of it (energy not growing with $N$), 
say, by keeping $\chi$
independent of $N$ and scaling $\Gamma_{0}$ by $N$, then 
one can easily see 
that this special transition point would behave like a true
quantum critical point 
with the gap vanishing as $1/\sqrt{N}$ \cite{Farhi-Grover}. 
The two states- the one localized at the hole and the other delocalized
over the entire lattice, have exactly equal energies at this ``critical" point
as the gap closes.
But before reaching that point 
(i.e., in the kinetic energy dominated region),
the ground state is non-degenerate and delocalized,
while after the point (i.e., in the potential energy dominated region)
the ground state is degenerate and localized at the hole.
Since there is no energy difference
between the two states (the localized and the delocalized ones)
at the critical point, the evolving system cannot sense
the global exchange of character between the ground state
and the first excited state at this point and thus fails to 
adopt (tunnel) accordingly to follow the ground state. 
%Right at this 
%point, the total energy can be minimized exactly to the same minimum value 
%in two different ways -
%either by delocalizing the wave-function
%over all the sites, 
%(which had been the more
%effective way to reduce energy before reaching the transition point, 
%since the kinetic would be dominating in that region) 
%or by localizing it just at the hole. 
%Thus at this point, there are two degenerate ground states
%with completely different global characters - one  
%localized at the hole, while the other delocalized
%over all the sites, with the energy gap vanishing between them.
%Before the critical point (i.e., in the kinetic energy dominated region),
%the ground state is non-degenerate and delocalized, 
%while after the point (i.e., in the potential energy dominated region) 
%the ground state is degenerate and localized at the hole.
%Since there is no energy difference
%between the ground state and the first excited state at the
%critical point, the evolving 
%wave-function cannot sense the global change in the character of 
%the ground state and fails to follow it. 
Letting the hole scale linearly
with $N$ and taking infinite range hopping, one prevents the gap from closing,
and thus favors the localized state energetically over the delocalized one
at the transition point.
This acts as a drive for the necessary tunneling and thus provides a guidance
for the system to follow the ground state as it 
changes its global character.
% unless it is infinitely slow in passing through
%the point.        
%(which would be the the more
%efficient way of reducing energy 
%after passing through the transition point, since potential energy dominates
%in there). 

However, it is also worth noting that increasing the hole-depth $\chi_{0}$ 
indiscriminately does not pay. 
In the case of constant-$\Gamma$ annealing, the adiabatic factor increases
with $r$ (Eq.\ref{ConstG-af-max}), which means we have to keep $r$ as small
as possible, i.e., $r \rightarrow 1$ gives the best result.
This can be 
explained by noting that a larger $\chi_{0}$ would demand a slower rate of
evolution if adiabaticity is to be ensured.
In the constant-$\chi$
case, adiabatic factor decreases with $r$, but
one has to keep $r < 1$, which means $\chi$ 
can be increased only linearly with $N$, up to the upper limit  $N\Gamma_{0}$
in order to accelerate the annealing effectively. 
This is because, too big $\chi_{0}$ implies too much error
in the initial state and the overlap is negligible right from the onset. 
Hence even with a perfectly adiabatic evolution 
we end up with the negligible overlap.

To summarize, we studied the adiabatic searching of
a gradient-free potential energy landscape on a lattice 
with a randomly placed single isolated minimum
(hole) by a quantum searcher. We have found that the depth of the hole plays 
a non-classical role in accelerating the search. This is because the 
delocalized wave-function of the quantum adiabatic searcher can detect the 
hole from the very onset of the search and keep track of it, in contrast to
a classical searcher, who of course cannot sense the hole depth non-locally
and utilize it. We found that the condition for adiabatic search requires
an infinite-range hopping between the points in the PEL and the depth of
the hole to scale linearly with the lattice-size $N$. We have discussed how
the hole-depth plays a role in preventing the closure of the gap 
and the consequent divergence of the searching time.

\vspace{0.7cm}

\noindent
\textbf{\leftline{Acknowledgments:}} \\
\noindent The author is thankful to B. K. Chakrabarti, G. E. Santoro and 
E. Tosatti for useful discussions. The author also gratefully acknowledges
the kind hospitality of J.-I. Inoue during the author's visit to Hokkaido University,
where the manuscript was completed.

\end{document}